\documentstyle[12pt,epsfig,subfigure]{article}
\title{PHOTON-PHOTON TOTAL INELASTIC CROSS-SECTION\footnote{Talk presented
 at Photon'97, Egmond aan Zee, May 1997}}
\author{A. CORSETTI\\
\smallskip
Physics Department, Northeastern University, Boston, USA\\
\medskip
R. M. GODBOLE\\
\smallskip
Center for Theoretical Studies, Indian Institute of Science, 
Bangalore, India\\
\medskip
and\\
G. PANCHERI\\
\smallskip
INFN - Laboratori Nazionali di Frascati, Frascati, Italy}
\begin{document}
\maketitle
\begin{abstract}
We discuss  predictions for  the total inelastic $\gamma \gamma$
cross-section and  their model dependence  on the input parameters. We compare  
results from a simple extension of the Regge Pomeron exchange model as well 
as  predictions from the  eikonalized mini-jet model  with recent LEP data.

\end{abstract}

It is by now established that all total cross-sections, including
photoproduction, rise as the c.m. energy of the colliding particles 
increases.  So far a successful description of total
cross-sections is obtained in the Regge/Pomeron exchange model
\cite{DL}, in which a Regge pole and a Pomeron are 
exchanged and   total cross-sections are seen to first decrease
 and subsequently rise
according to the expression

$$\sigma^{tot}_{ab}=Y_{ab} s^{-\eta}+X_{ab}s^{\epsilon}$$
where $\epsilon $ and $\eta$ are related to the intercept at zero of the
leading Regge trajectory and of the Pomeron, respectively
 $\eta\approx 0.5$ and $\epsilon \approx 0.08$. This parametrization applies
successfully \cite{DL} to  photoproduction, as shown in Fig. 1,
and  to the lower
energy data on $\gamma \gamma$\cite{SJOS}.
Assuming the hypothesis of factorization at the poles, one can make a
prediction for 
$\gamma \gamma$ total inelastic cross-section, using
 $$Y_{ab}^2=Y_{aa} Y_{bb}\ \ \ \ \ \ X_{ab}^2=X_{aa} X_{bb}$$
and    extracting the coefficients X and Y from those for  the fit to
 photo-production and hadron-hadron data. 
In particular, using for
$\eta$ and $\epsilon$ the average values from the Particle Data Group
compilation \cite{PDG} and averaging among the $pp$ and $\bar p p$
coefficients, one can have a first check of the factorization hypothesis. 
Noticing that the coefficient Y from photoproduction data has a
large error and that   prediction from the Regge/Pomeron exchange model 
refer to total cross-sections rather than the inelastic ones, 
these predictions can be enlarged into a band as shown  in Fig.2.
 
An alternative model for the rise of all total cross-sections,
relies on hard parton-parton scattering. It was suggested \cite{CLINE}
that hard 
collisions between  elementary constituents of the colliding hadrons,
the partons,  could be responsible for this rise which starts around  
$\sqrt{s} \ge 10\div 20 \ GeV$.   This suggestion has subsequently evolved 
into mini-jet models \cite{minijet}, whose eikonal formulation
satisfies unitarity 
while embodying the  concepts of rising total cross-sections with rising 
jet cross-sections. For processes involving photons,  the model has to 
incorporate \cite{ladinsky} the hadronisation probability $P_\gamma^{had}$
for the photon to fluctuate itself into a hadronic state. The eikonalised 
mini--jet cross-section is then   
\begin{equation}
\label{eikonal}
\sigma^{inel}_{ab} = P^{had}_{ab}\int d^2\vec{b}[1-e^{n(b,s)}]
\end{equation}
with the average number of collisions at a given impact
parameter $\vec{b}$ given by  
\begin{equation}
\label{av_n}
n(b,s)=A_{ab} (b) (\sigma^{soft}_{ab} + {{1}\over{P^{had}_{ab}}}
\sigma^{jet}_{ab})
\end{equation} 
In eqs.(\ref{eikonal}, \ref{av_n}),
 $P^{had}_{ab}$ is the probability that  the colliding particles
$a,b$ are both in a hadronic state,  
$A_{ab} (b)$ describes the transverse overlap of the  partons 
in the two projectiles  normalised to 1,
$\sigma^{soft}_{ab}$ is the non-perturbative part of the cross-section from 
which the factor of $P_{ab}^{had} $ has already been factored out and 
$\sigma^{jet}_{ab} $ is the hard part of the cross--section. The basic statement
of the mini-jet model for total cross-sections is  that the rise in 
$\sigma^{jet}_{ab} $ drives the rise of $\sigma_{ab}^{inel}$ with energy. 
Letting
\begin{equation}
\label{phad}
P_{\gamma p}^{had} = P_{\gamma}^{had} \ \ \ \ \ and \ \ \ 
 P_{\gamma \gamma}^{had} \approx   (P_{\gamma}^{had})^2
\end{equation}
one can   extrapolate the model from photoproduction to photon-photon 
collisions. The issue of total $\gamma \gamma$ cross-sections assumes an additional
significance in view of the large potential backgrounds  that Beamstrahlung
photons could cause at future Linear Colliders
 \cite{messy}. Because the hadronic structure of the photon involves 
both a perturbative and nonperturbative part, it has been  
proposed~\cite{SJOS,SARC} to use  a sum of eikonalized functions instead of
eq.(\ref{eikonal}) in processes  involving photons.

The  predictions of the eikonalised mini-jet model for photon induced processes
thus depend on 1) the assumption of one or more eikonals 2)  the hard jet 
cross-section 
$\sigma_{jet}=\int_{p_{tmin}} {{d^2\hat{\sigma}}\over{dp_t^2}} dp_t^2$ 
which in turn depends on the minimum 
$p_t$ above which one can expect perturbative QCD to hold viz. $ p_{tmin}$
and the  parton densities in the colliding particles $a$ and $b$, 
3) the soft cross--section $\sigma^{soft}_{ab}$ 4) the overlap function
$ A_{ab}(b) $, defined as 
\begin{equation}
\label{aob}
A(b)={{1}\over{(2\pi)^2}}\int d^2\vec{q}{\cal F}_1(q) {\cal F}_2(q) 
e^{i\vec{q}\cdot \vec{b}}
\end{equation}
 where ${\cal F}$ is the Fourier transform of the b-distribution
of partons in the colliding particles and 5) last, but not the least, 
$P_{ab}^{had}$.

In this note we shall restrict ourselves to a single eikonal. The hard jet
cross-sections are calculated in LO perturbative QCD and  use 
 photonic parton densities  GRV \cite{GRV} calculated to the
  leading order. We  
determine $\sigma_{\gamma \gamma}^{soft}$ from $\sigma_{\gamma p}^{soft}$ 
which in turn is  determined by a fit to the photoproduction data.
From inspection of the photoproduction  data, one
can assume that $\sigma_{soft}$ should contain both a constant and an 
energy decreasing term. Following the suggestion\cite{SARC}
\begin{equation}
\label{soft}
\sigma^{soft}_{\gamma p} =\sigma^0 +
{{A}\over{\sqrt{s}}}+{{B}\over{s}}
\end{equation}
we then calculate values for $\sigma^0, A$ and $B$ from a best fit \cite{thesis}
to the 
low energy photoproduction data, starting with the Quark Parton Model ansatz
$\sigma^0_{\gamma p}\approx {{2}\over{3}}\sigma^0_{pp}$. For $\gamma \gamma$
collisions, we  repeat the QPM suggestion and propose
\begin{equation}
\sigma^{soft}_{\gamma \gamma}={{2}\over{3}} \sigma^{soft}_{\gamma p},\ i.e.\ 
\sigma^0_{\gamma\gamma}=20.8 mb,A_{\gamma \gamma}=6.7\ mb\ GeV^{3/2},
B_{\gamma \gamma}= 25.3 \ mb\ GeV
\end{equation}
Whereas the effect of the uncertainties in the above three quantities on the
predictions of the inelastic photoproduction and $\gamma \gamma$ cross-sections
has been studied in literature to a fair extent 
\cite{SJOS,SARC,FS} the effect of the other two has not been 
much discussed.
In the original use of the eikonal model, the overlap function $A_{ab} (b)$ of
eq.(\ref{aob})  is obtained using for ${\cal F}$ the electromagnetic form 
factors. For protons this is given by the dipole expression
\begin{equation}
\label{dipole}
{\cal F}_{prot}(q)=[{{\nu^2}\over{q^2+\nu^2}}]^2
\end{equation}
with $\nu^2=0.71\ GeV^2$. For photons a number of authors \cite{SARC,FLETCHER}, 
 on the basis of Vector Meson Dominance, have assumed the same functional 
form as for pion, i.e. the pole expression 
\begin{equation}
\label{pole}
{\cal F}_{pion}(q)={{k_0^2}\over{q^2+k_0^2}}\ \ \ with \ \ \ k_0=0.735\ GeV.
\end{equation}
 There also exists  
another possibility, i.e. that the b-space 
distribution of partons is the Fourier transform of their intrinsic 
transverse momentum distributions~\cite{BN}. While 
for the proton this would correspond to use a Gaussian 
distribution instead of the dipole expression, eq.(\ref{dipole}),
for the photon one can argue that the intrinsic transverse momentum ansatz
\cite{rohini} would imply the use of a different value of the parameter 
$k_o$\cite{ZEUS} in the pole expression for the form factor.
By varying $k_o$ one can then explore  both the intrinsic transverse 
distribution case and the form factor cum VMD hypothesis. Notice that the 
region most important to this calculation is for large values of the parameter 
b, where the overlap function changes trend, and is larger for smaller 
$k_o$ values.

Let us now look at $P^{had}_\gamma$.  This is clearly expected to be 
${\cal O} (\alpha_{em})$. Based on  Vector Meson Dominance one expects,
\begin{equation}
\label{PVMD}
P_\gamma^{had} = P_{VMD}=\sum_{V=\rho,\omega,\phi} {{4\pi \alpha}
\over{f^2_V}}= {{1}\over{250}}
\end{equation}
Although in principle, $P^{had}_\gamma$ is not a
constant,  for simplicity, we adopt here a fixed
value\cite{FLETCHER} of 
  1/204, which includes a non-VMD contribution of $\approx 20\%$.
  Notice that a fixed value of $P_{had}$ can be absorbed into a 
  redefinition of the parameter $k_o$ through a simple change of variables
  \cite{review}.

Having thus established the range of variability of the quantities 
involved in the calculation of
 total inelastic photonic cross sections, we can proceed to  
 compare the predictions of the eikonalized minijet model with data.
  We use  GRV (LO)  densities
 and  show  the mini-jet result in Fig.1,
 using the form factor model for A(b),
i.e. eq.(\ref{aob}) with $k_o=0.735\ GeV$.
 In the figures, we have not added
the direct contribution, which will slightly increase the cross-section 
in the 10 GeV region. We observe that it is possible to include the high 
energy points using GRV densities and $p_{tmin}=2\ GeV$, but the low energy 
region would be better described  by a smaller $p_{tmin}$. This is the region 
where the rise, according to some authors, notably within the framework of 
the Dual Parton Model, is attributed to the  so-called
{\it soft Pomeron}. 

\begin{figure}
\centering
\mbox{\subfigure
[Fig.1: Total 
inelastic photon-proton cross-section]
{\epsfig{figure=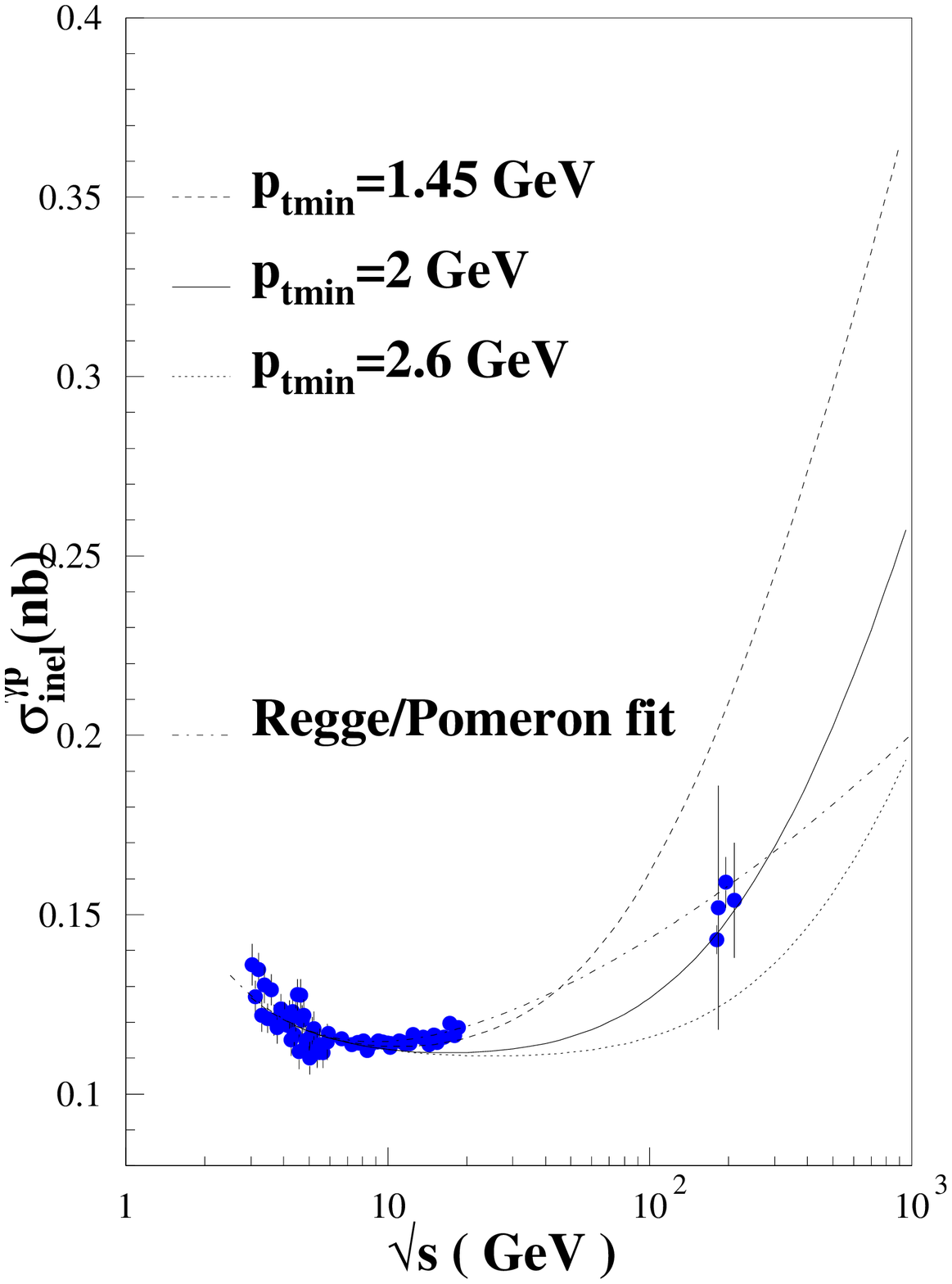,width=0.5\textwidth}}\quad
\subfigure
[Fig.2: Total inelastic photon-photon cross-section.]
{\epsfig{figure=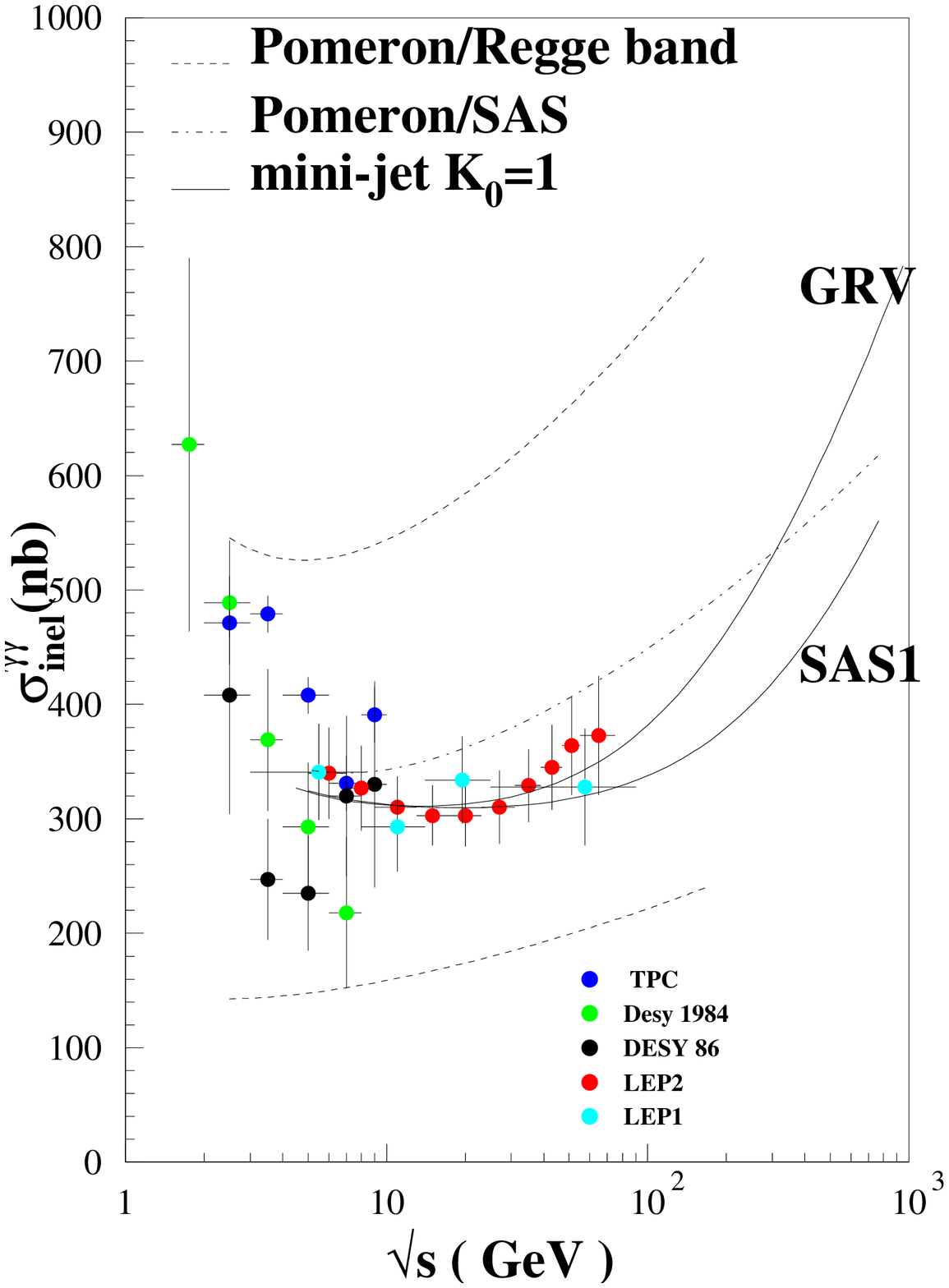,width=.5\textwidth}}}
\end{figure}



We now apply the same criteria and parameter set used in
 $\gamma p$ collisions to the case of photon-photon collisions,
i.e. $P_{h/\gamma}=1/204$, $p_{tmin}=2\ GeV$ 
and A(b) from eq.(\ref{aob}).
 A comparison
with  $\gamma \gamma$ data shows that although the value $k_o=0.735$,
corresponding to the pion-factor, is
 compatible with the low energy data up to $10\ GeV$ \cite{desy96} within
 the limits established by the large errors involved,
 at higher energies \cite{thisconf}
the best fit is obtained using a slightly larger value, i.e. $k_0=1\ GeV$,
and this is the one used in Fig.2. For comparison, we have also added
mini-jet model predictions with SAS1 photon densities \cite{SAS1}and
predictions (Pomeron/SaS) 
based on a Pomeron/Regge type parametrization\cite{SJOS}.
 


\begin{thebibliography}{99}
\bibitem{DL} A. Donnachie and P.V. Landshoff, Phys. Lett. B296 (1992) 227.
\label{DL}
\bibitem{SJOS}
G. Schuler and T. Sj\"ostrand, Phys. Lett. {\bf B 300} (1993) 169, Nucl. Phys. 
{\bf B 407} (1993) 539, CERN-TH/95-62.
\label{SJOS}
\bibitem{PDG}
 Particle Data Group, Physical Review D54 (1996) 191.
\label{PDG}
\bibitem{CLINE}
D.Cline, F.Halzen and J. Luthe, Phys. Rev. Lett. {\bf 31} (1973) 491.
\label{CLINE}
\bibitem{minijet}
A. Capella and J. Tran Thanh Van, Z. Phys. {\bf C23} (1984)168,
T.Gaisser and F.Halzen, Phys. Rev. Lett. {\bf54}  (1985) 1754,
G.Pancheri and Y.N.Srivastava, Phys. Letters {\bf B182} (1985),
P. l`Heureux, B. Margolis and P. Valin, Phys. Rev. {\bf D 32} (1985) 1681,
L. Durand and H. Pi, Phys. Rev. Lett. {\bf 58} (1987) 58.
\label{minijets}
\bibitem{ladinsky}
J.C. Collins and G.A. Ladinsky, Phys. Rev. {\bf D 43} (1991) 2847.
\label{ladinsky}
\bibitem{messy}
M. Drees and R.M. Godbole, Phys. Rev. Lett. {\bf 67} (1991) 1189,
P. Chen, T.L. Barklow and M. E. Peskin, Phys. Rev. {\bf D 49} (1994) 3209.
\label{messy} 

\bibitem{SARC}
K. Honjo,  L. Durand, R. Gandhi, H. Pi and I. Sarcevic, Phys. Rev. {\bf D 48} 
(1993) 1048. 
\label{SARC}

\bibitem{GRV}
M. Gl\"uck, E. Reya and A. Vogt, Phys. Rev. {\bf D 46} (1992) 1973.
\label{GRV}
\bibitem{thesis} A. Corsetti, September 1994 Laurea Thesis, University of Rome
La Sapienza.
\label{thesis}
\bibitem{FS}
J.R. Foreshaw and J.K. Storrow, Phys. Lett. {\bf B 278} (1992) 193; Phys. Rev. 
{\bf D46} (1992) 3279.
\label{FS}
\bibitem{FLETCHER}
R.S. Fletcher , T.K. Gaisser and F.Halzen, Phys. Rev. {\bf D 45} (1992) 377; 
erratum Phys. Rev. {\bf D 45} (1992) 3279.
\label{FLETCHER}
\bibitem{BN} A. Corsetti,  Grau, G. Pancheri and Y.N. Srivastava, 
{\bf PLB 382}
(1996) 282.
\label{BN}
\bibitem{rohini} J. Field, E. Pietarinen and K. Kajantie, Nucl. Phys. 
{\bf B 171} (1980) 377; M. Drees, {\it Proceedings of 23rd International 
Symposium on Multiparticle Dynamics}, Aspen, Colo., Sep. 1993. Eds. M.M. Block
and A.R. White. 
\bibitem{ZEUS}
M. Derrick et al., ZEUS collaboration, PLB 354 (1995) 163.
\label{ZEUS}
\bibitem{review}
M. Drees, Univ. Wisconsin report {\bf MAD/PH-95-867},{\it  Proceedings of the
4th workshop on TRISTAN physics at High Luminosities}, 
KEK, Tsukuba, Japan, Nov. 1994;
M. Drees and R. Godbole, J. Phys. G 21 (1995) 1559.
\label{review}
\bibitem{desy96}A. Corsetti, R. Godbole and G. Pancheri, in
Proceedings of the Workshop on $e^+e^- $ Collisions at TeV Energies, 
Annecy, Gran Sasso, Hamburg, DESY 96-123D, June 1996, page. 495.
\label{desy96}
\bibitem{thisconf}
W. van Rossum,  these propceedings.
\label{thisconf}
\bibitem{SAS1}
G. Shuler and T. Sjostrand, Zeit Phys. C68 (1995) 607: Phys., Lett. B376
(1996) 193.

\end{thebibliography}
\end{document}